\documentclass[twocolumn,showpacs,preprintnumbers,amsmath,amssymb]{revtex4}

\usepackage{graphicx}
\usepackage{dcolumn}
\usepackage{bm}
\usepackage{epstopdf}

\begin{document}

\preprint{}

\title{Accelerator-based Single-shot Ultrafast Transmission Electron Microscope with Picosecond Temporal Resolution and Nanometer Spatial Resolution}
\author{D. Xiang\footnote{dxiang@sjtu.edu.cn}, F. Fu and J. Zhang}
\affiliation{
Key Laboratory for Laser Plasmas (Ministry of Education), Department of Physics and Astronomy, Shanghai Jiao Tong University, Shanghai 200240, China}

\author{X. Huang, L. Wang and X. Wang}
\affiliation{SLAC National Accelerator Laboratory, Menlo Park, CA, 94025, USA
}
\author{W. Wan}
\affiliation{Lawrence Berkeley National Laboratory, Berkeley, CA, 94720, USA
}
\date{\today}

\begin{abstract}

We present feasibility study of an accelerator-based ultrafast transmission electron microscope (u-TEM) capable of producing a full field image in a single-shot with simultaneous picosecond temporal resolution and nanometer spatial resolution. We study key physics related to performance of u-TEMs, and discuss major challenges as well as possible solutions for practical realization of u-TEMs. The feasibility of u-TEMs is confirmed through simulations using realistic electron beam parameters. We anticipate that u-TEMs with a product of temporal and spatial resolution beyond $10^{-19}~$m*s will open up new opportunities in probing matter at ultrafast temporal and ultrasmall spatial scales.

\end{abstract}

\maketitle

\section{Introduction}
Transmission electron microscope (TEM \cite{tem1}) has played an important role in development of physics, chemistry, biology and material sciences. In a conventional TEM (see, for example \cite{tembook1, tembook2}), the electrons are produced in an electron gun, accelerated with DC field, and then focused onto a sample with condenser lens system. The electron beam distribution at the sample is further magnified (by up to a few millions) with the imaging lens system, and finally measured with an area detector. Over several decades TEMs have been widely used to probe molecules, atoms, crystals, and innovative new materials with atomic resolution in order to better understand their properties and behaviors. Some breakthroughs enabled by TEMs include discovery of the single-walled nanotube \cite{nanotube}, detection of tumor viruses \cite{virus}, imaging of individual atoms \cite{atom}, just to name a few.

In conventional full field TEMs, the electron beam is typically produced with thermionic, Schottky, or field emission. The voltage of routine TEMs is below 200 kV and medium-voltage TEMs work at 200-500 kV to provide better resolution. In high-voltage electron microscopes, the voltage reaches 500 kV to 3 MV \cite{3MeV}, which provides much better transmission such that materials and large biological cells that are difficult to prepare in thin enough layers can be studied. As the electrons are emitted continuously, the temporal resolution in conventional TEMs is only achieved at the millisecond level using a fast framing camera. Though millisecond is not fast, it has already allowed real-time study of some slow dynamics, e.g. the motion of thermally activated vortices in a superconductor has been observed with an electron microscope at 30 frames per second \cite{Naturesc}. The temporal resolution can be significantly improved if the electrons are bunched (e.g. illuminating a photocathode with a short pulse laser to produce a short electron beam) rather than emitted in constant stream. This also removes the need for a fast detector and the temporal resolution is simply determined by the electron bunch length, which in many cases is comparable to the laser pulse width.

Currently, there are two major configurations for achieving high temporal resolution in TEMs. The first configuration operates in stroboscopic mode \cite{Zewail} in which a femtosecond beam with only a single electron (on average) to avoid space-charge effect is used to probe the sample after a femotsecond pump laser. Typically one useful image corresponding to a specific time delay between the pump laser and probe electron beam is obtained with integration over about $10^8$ shots. While very high temporal resolution and spatial resolution can be achieved with this configuration, it only applies to studies of perfectly reversible process, because the sample needs to be pumped $\sim10^8$ times and the sample must completely recover after each shot. Alternatively, a useful image may be obtained in a single shot with a longer pulse that contains enough electrons. With the beam peak current several orders of magnitude higher than a conventional TEM, the temporal resolution and spatial resolution in this configuration is degraded by space charge effects and the limited electron beam brightness, etc. For instance, the recently developed dynamic TEM (DTEM) has achieved about 15 nanosecond (ns) temporal resolution and 10 nanometer (nm) spatial resolution (corresponding to a product of temporal and spatial resolution $10^{-16}~$m*s) using a 200 kV electron beam produced with a nanosecond laser pulse \cite{DTEM}.

Increasing both the beam energy and beam brightness may further push the product of temporal and spatial resolution by a few orders of magnitude (see, for example \cite{XJLDRD}). For instance, a possible path to reach 10 ps temporal resolution and 10 nm spatial resolution with a 5 MV TEM has been briefly discussed in \cite{APL10ps}. A prototype u-TEM using an electron beam produced in a photocathode rf gun has been developed and about 300 nm spatial resolution has been achieved \cite{Yang1, Yang2}. Very recently, an u-TEM using quadrupoles for imaging has been briefly discussed in \cite{UEMUCLA}. In this paper, we present a detailed study of the feasibility of an accelerator-based u-TEM capable of providing a few picosecond temporal resolution and a few nanometer spatial resolution in a single-shot. We study key physics related to performance of u-TEMs, and discuss major challenges (such as beam emittance, beam energy spread, beam energy stability, space charge effect, etc.) as well as possible solutions for practical realization of u-TEMs. Using a representative set of parameters, the feasibility of achieving a product of temporal and spatial resolution beyond $10^{-19}~$m*s is confirmed through simulation.

\section{Considerations for an u-TEM}
The performance of a conventional TEM is mainly determined by the electron beam quality and the spherical and chromatic aberrations of the imaging system. The spatial resolution limited by spherical aberration is approximately $r_s=C_s\theta^3$, where $C_s$ is the spherical aberration coefficient and $\theta$ is the divergence of the detected electrons at the sample. Similarly, the spatial resolution limited by chromatical aberration is approximately $r_c=C_c\theta\delta$, where $C_c$ is the chromatical aberration coefficient and $\delta$ is the relative energy spread of the beam at the exit of the sample. Both $C_s$ and $C_c$ are on the order of the focal length of the objective lens. In accelerator terminology, $C_s$ and $C_c$ are related to the $U_{1222}$ and $T_{126}$ elements of the third order and second order transfer matrix, respectively. The biggest difference between an u-TEM and a conventional TEM is, perhaps, that the beam peak current is many orders of magnitude higher such that collective self-interactions of the electrons may play a role. In this section we discuss the general considerations for a single-shot u-TEM and give an order-of-magnitude estimate for the beam and imaging system requirements for realization of an u-TEM.

In view of producing a beam with high brightness, a high accelerating field gradient is required, regardless of if the electrons are produced in a DC diode or a laser-driven photocathode rf gun. When electrons are produced, they are affected by their self-fields. For sufficiently high current density, the electric field at the cathode from space charge may equal to the external field, and the current density can not be increased further. For a planar diode with gap $d$ and voltage $V$, the maximal current density limited by space charge is proportional to $V^{3/2}/d^2$ \cite{Reiser}. For a laser-driven photocathode rf gun, the maximal current density is proportional to the external field. Therefore, higher accelerating field gradient allows extraction of a given charge from a source with smaller area that reduces thermal emittance (given the same transverse thermal energy at the cathode) and increases the space charge limited maximal beam brightness \cite{Bazarov}. Furthermore, higher gradient also allows electrons to be accelerated to relativistic within a shorter distance that mitigates space charge induced emittance growth, which is useful to preserve the beam brightness.

In view of mitigating collective self-interaction of the electrons that may change electron trajectory and energy in the imaging system to degrade the spatial resolution, a high beam energy is required. In general, the transverse space charge force that changes particle's trajectory is proportional to $1/\gamma^{2}$ due to the cancelation of electric and magnetic forces, where $\gamma$ is the relativistic factor of the beam (see, for example \cite{Chao}). The longitudinal space charge force that changes particle's energy is proportional to $I'(z)/\gamma^2$, where $I'(z)$ is the derivative of the beam current at a longitudinal position $z$. Therefore, a beam with high energy and flat current distribution is desired for mitigating longitudinal space charge force that may increase beam energy spread. Note, for a coasting beam as in conventional TEMs, the electric and magnetic fields are purely transverse by symmetry. So longitudinal space charge is not an issue of concern for conventional TEMs and the transverse space charge force acts like a weak defocusing lens and can be readily compensated by increasing the strength of the solenoids.

In view of reducing spherical and chromatic aberrations of the imaging system, solenoids with strong strengths to provide a short focal length (thus smaller spherical and chromatic aberration coefficients) are needed. This will also loosen the requirements on beam emittance, energy spread and energy stability.

Based on these considerations, the proposed u-TEM is based on accelerator that provides a much higher accelerating field than a conventional TEM. Similar to conventional TEMs, an u-TEM consists of three main systems: the electron source, condenser lens system, and imaging system, as shown schematically in Fig.~1. To provide picosecond temporal resolution, the electron beam is produced in a photocathode rf gun illuminated with picosecond laser pulse. The beam energy at the exit of the gun is typically a few MeV, which effectively mitigates the temporal broadening from space charge effect and makes it possible to preserve the picosecond pulse width during transport to the sample. In our representative design the electron beam is produced in a standard S-band photocathode rf gun (with frequency at 2.856 GHz), and the beam global energy spread is further reduced in a harmonic rf cavity (e.g. C-band cavity with frequency at 5.712 GHz). A solenoid (S) following the gun is used to minimize emittance growth from space charge effect and to control the beam size. The condenser lens (C) used to focus the beam onto a sample and imaging system, composed of an objective lens (O), an intermediate lens (I) and a projection lens (P), that magnifies the beam, play similar roles as in conventional u-TEM, except that their strengths are much higher to provide sufficient focusing for MeV beam (depending on beam energy and magnetic field strength, superconducting solenoids may be used). Finally the magnified beam is measured with an area detector (D) (e.g. a fluorescent screen and a CCD camera).
   \begin{figure}[t]
       \includegraphics[width = 0.49\textwidth]{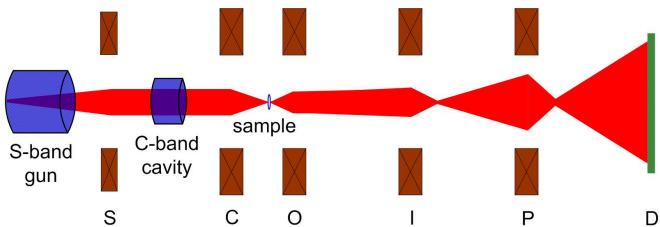}
    \caption{Schematic of an accelerator-based u-TEM. It consists of a photocathode rf gun to produce picosecond electron beam, a condenser lens (C) to focus the electron beam onto the sample, an objective lens (O), an intermediate lens (I) and an projection lens (P) to magnify the electron beam, and an area detector (D) to measure the magnified beam distribution.
    \label{beamforuem}}
    \end{figure}

\subsection{Electron source requirement}
In an optical microscope the image contrast is produced by the variation in optical absorption on the specimen. However, for TEM with the sample thickness comparable to the mean free path, electrons are not absorbed by the specimen. Instead, the contrast is produced by blocking those electrons whose angle after scattering is larger than the acceptance of objective aperture located at the back focal plane of the objective lens (bright-field imaging mode). Alternatively, one may block the electrons that pass through the sample without scattering and only use those with large angles for imaging, as in the dark-field imaging mode. In either case, the beam intrinsic divergence should be smaller than the scattering angle for optimal performance. Assuming the sample thickness is comparable to mean free path, elastic scattering is the dominant process, and the characteristic scattering angle is $\theta_0=\lambda/2\pi r$ (see, for example \cite{tembook1, tembook2}), where $\lambda$ is the De Broglie wavelength of the electron, $r=a_H Z^{-1/3}$ is the screening radius, $a_H=52.9~$pm is the Bohr radius and $Z$ is the atomic number of the material. For a representative relativistic electron beam with 4 MeV kinetic energy the wavelength is about 0.27 pm, and the corresponding characteristic scattering angle is a few mrad (e.g. $\theta_0\approx1.5~$ mrad for low-$Z$ material Carbon with $Z=6$ and $\theta_0\approx3.6~$ mrad for high-$Z$ material Platinum with $Z=78$).

It is then straightforward to have an order-of-magnitude estimate of the emittance requirement for single-shot u-TEMs. Assuming the rms beam size at the sample is 1 $\mu$m, the normalized emittance for a beam with 4 MeV kinetic energy should be around 10 nm to make the rms divergence (about 1 mrad) smaller than the characteristic scattering angle. Assuming half of the electrons are scattered and stopped by the aperture of the objective lens, a beam charge of 1.6 pC is needed to achieve sufficient signal-to-noise ratio with 10 nm spatial resolution (corresponding to approximately 100 detected electrons per 100 nm$^2$ as required by the Rose criterion \cite{Rose}).

Assuming $C_s\approx C_c\approx2~$ cm (without any aberration correction elements), an energy spread on the order of $10^{-4}$ is required to achieve a few nanometer resolution (the acceptance angle of the aperture is taken as 2 mrad). As we will show later, such a beam (1.6 pC with 10 nm normalized emittance and $10^{-4}$ energy spread) is within reach with the state-of-the-art photocathode rf guns. It should be noted that world's highest voltage TEM with 3 MeV beam stands 13 m high and weighs 130 tons \cite{3MeV} (the acceleration tube alone stands 6 m high). With the electron beam produced in a photocathode rf gun, the u-TEM with a few MeV beam can be also made much more compact (e.g. the length of an S-band gun is only $\sim$15 cm).

\subsection{Imaging system requirement}
The core element of a TEM imaging system is the solenoid. As we will show later many of the physics related to TEMs can be understood by analyzing a simple imaging system. For simplicity, here we study an imaging system that consists of a hard-edge solenoid with length $L$ and focusing strength $K$ preceded by a drift with length $L_1$ and followed by another drift with length $L_2$. The first order transfer matrix for such a system with the beam coordinates rotated by $-KL$ (to decouple the beam dynamics in $x$ and $y$) can be easily found (see, for example \cite{handbook}),
\begin{equation}
R=
\left[
\begin{array}{cc}
C-KSL_2 &  L_1(C-KSL_2)+(CL_2+S/K)  \\
-KS &  C-KSL_1   \\
\end{array}
\right],
\end{equation}
where $K=B_0/2B\rho$ with $B_0$ being the field inside the solenoid, $B\rho$ being the momentum of the central trajectory, $C=\cos(KL)$ and $S=\sin(KL)$. The focal length of the solenoid is $f=-1/R_{21}=1/KS$ (positive for focusing with this notation).

Point-to-point imaging is achieved when $R_{12}=0$, i.e.
\begin{equation}
L_1=\frac{KL_2C+S}{K(KL_2S-C)},
\end{equation}
with the magnification being $M=R_{11}\approx L_2/f$ assuming $L_2$ is much larger than the focal length. Under the imaging condition, Eq.~(1) is simplified to,
\begin{equation}
R=
\left[
\begin{array}{cc}
M &  0  \\
-1/f &  1/M   \\
\end{array}
\right],
\end{equation}

For the imaging lens system in a conventional TEM that consists of an objective lens, an intermediate lens and a projection lens, the subsequent lens uses the image of the upstream lens as the object and by cascading multiple stages the magnification of the imaging system can be made as large as a few millions.

In practical conditions $R_{12}$ may not be negligibly small, due to slight variation of beam energy, focusing strength of the solenoid, vibration of the sample, etc. The propagation of this error can be studied by multiplication of the transfer matrices of each imaging lens. Consider an imaging system with first order transfer matrix $R^{(1)}$ followed by another imaging system with transfer matrix $R^{(2)}$, e.g.
\begin{equation}
R^{(1)}=
\left[
\begin{array}{cc}
M_1 &  \zeta_1  \\
-1/f_1 &  1/M_1   \\
\end{array}
\right],
R^{(2)}=
\left[
\begin{array}{cc}
M_2 &  \zeta_2  \\
-1/f_2 &  1/M_2   \\
\end{array}
\right],
\end{equation}
where $M_{1(2)}$, $f_{1(2)}$, and $\zeta_{1(2)}$ are the magnification, focal length and residual angle-to-position element of the first(second) imaging system, respectively. In writing Eq.~(4) we have assumed that $\zeta_{1(2)}$ has very small value such that the magnification and focal length are still very close to the ideal values. Note, omission of the small perturbations to the magnification and the focal length makes the matrices in Eq.~(4) non-symplectic, but the physics are essentially unaffected. The total transfer matrix for this two-stage imaging system is,
\begin{eqnarray}
&&R=R^{(2)}R^{(1)}
\nonumber\\
    &&
    =
\left[
\begin{array}{cc}
M_1M_2-\zeta_2/f_1 &  M_2\zeta_1+\zeta_2/M_1  \\
-(M_1/f_2+1/f_1M_2) &  1/M_1M_2-\zeta_1/f_2   \\
\end{array}
\right].
\end{eqnarray}

From Eq.~(5) we have $R_{12}=M_2\zeta_1+\zeta_2/M_1$, which indicates that the residual angular-to-position element in the first imaging system is amplified (by $M_2$) while that in the second imaging system is demagnified (by $M_1$). This suggests that the first imaging system (i.e. the imaging system with objective lens) is the most critical element in a TEM. Any mismatch of beam energy, focusing strength of the objective lens, and location of the sample that leads to imaging errors will be further amplified by the subsequent intermediate lens and projection lens systems.

Similarly, the propagation of high order aberrations can be studied by multiplication of the high order transfer matrix. In this paper we limit our analysis to second order, as the small beam divergence and energy spread in general make the third order and even higher order effects negligible. Similar to Eq.~(5), the total second order transfer matrix for an imaging system with first order transfer matrix $R^{(1)}$ and second order transfer matrix $T^{(1)}$ followed by another imaging system with first order transfer matrix $R^{(2)}$ and second order transfer matrix $T^{(2)}$ can be calculated as,
\begin{equation}
T_{ijk}=\sum_{m=1}^{6} R_{im}^{(2)} T_{mjk}^{(1)} +\sum_{m,n=1}^{6} T_{imn}^{(2)}R_{mj}^{(1)}R_{nk}^{(1)}
\end{equation}
For above studied two-stage imaging systems (assuming imaging condition is satisfied), the matrix element that characterizes the chromatic aberration is found to be,
\begin{equation}
T_{126}=M_2T_{126}^{(1)} + T_{126}^{(2)}/M_1
\end{equation}
So we see that the total $T_{126}$ is approximately the $T_{126}^{(1)}$ multiplied by the magnification ratio of the second system, similar to the propagation of the imaging errors. This indicates that in a TEM, the aberration is dominated by the first lens. This is why the objective lens, that immediately follows the specimen, is the most important element of a TEM. The physics behind Eq.~(5) and Eq.~(7) is that the angular magnification is inversely proportional to the lateral magnification, such that after magnification in the objective lens the angles of the electrons in all subsequent lenses are so small that it will not cause considerable degradation to the image.

Now let us take a look at the chromatic aberration of the imaging system used in Eq.~(1). When the imaging condition is met, the $T_{126}$ element can be found with Eq.~(6),
\begin{equation}
T_{126}=T_{116}^{(s)}L_1+T_{126}^{(s)}+L_2[T_{216}^{(s)}L_1+T_{226}^{(s)}]
\end{equation}
with
\begin{eqnarray}
&&T_{116}^{(s)}=T_{226}^{(s)}=\frac{1}{2}KL\sin(KL),
    \nonumber\\
    &&
T_{126}^{(s)}=\frac{\sin(KL)}{2K} - \frac{L}{2}\cos(KL),
    \nonumber\\
    &&
T_{216}^{(s)}=\frac{1}{2}K\big[ KL \cos(KL)+ \sin(KL) \big]
   .
\end{eqnarray}
Derivation of the second order transfer matrix element $T_{ijk}^{(s)}$ for a solenoid is given in the Appendix. It is straightforward to see that $T_{126}$ is dominated by $L_2T_{226}^{(s)}$. The chromatic aberration coefficient is thus,
\begin{equation}
C_c=2T_{126}/M \approx L
\end{equation}
Note, for a given magnification, the drift $L_2$ (and thus the space needed for a microscope) scales with $f$. To make a compact microscope, the length and strength of the solenoid is typically chosen to maximize $K\sin{KL}$, which yields approximately $KL\approx\pi/2$ and thus $C_c\approx \pi/2K\sim f$, a well-known result in electron microscope community that the chromatic aberration is on the order of the focal length of the objective lens. Therefore, the strength of the objective lens should be made as large as practically possible to reduce aberrations.
    \begin{figure*}[t]
    \includegraphics[width = 0.22\textwidth]{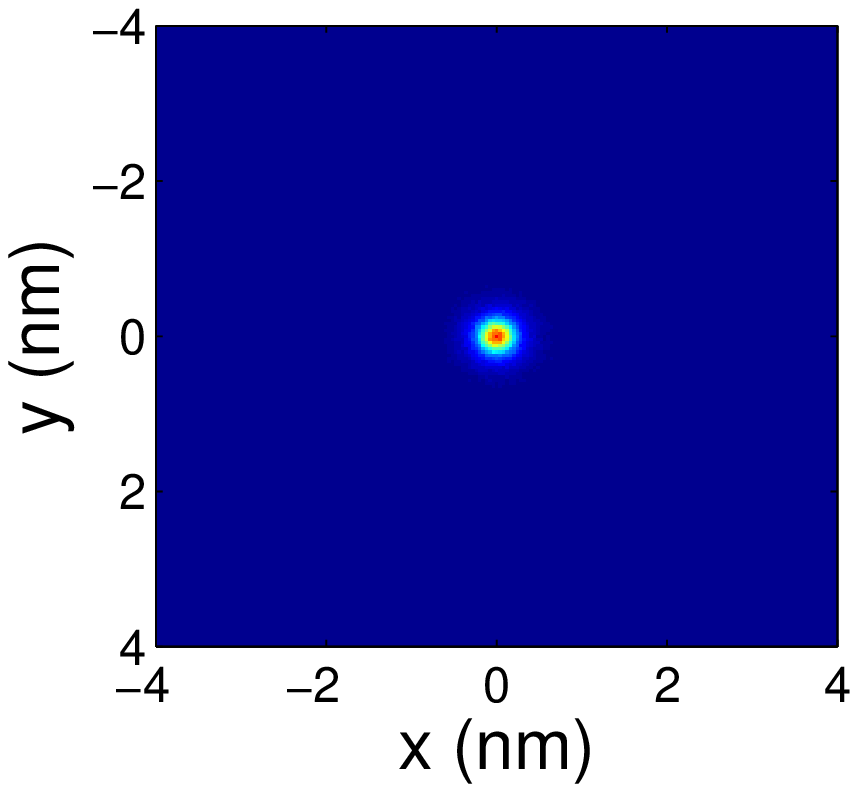}
     \includegraphics[width = 0.22\textwidth]{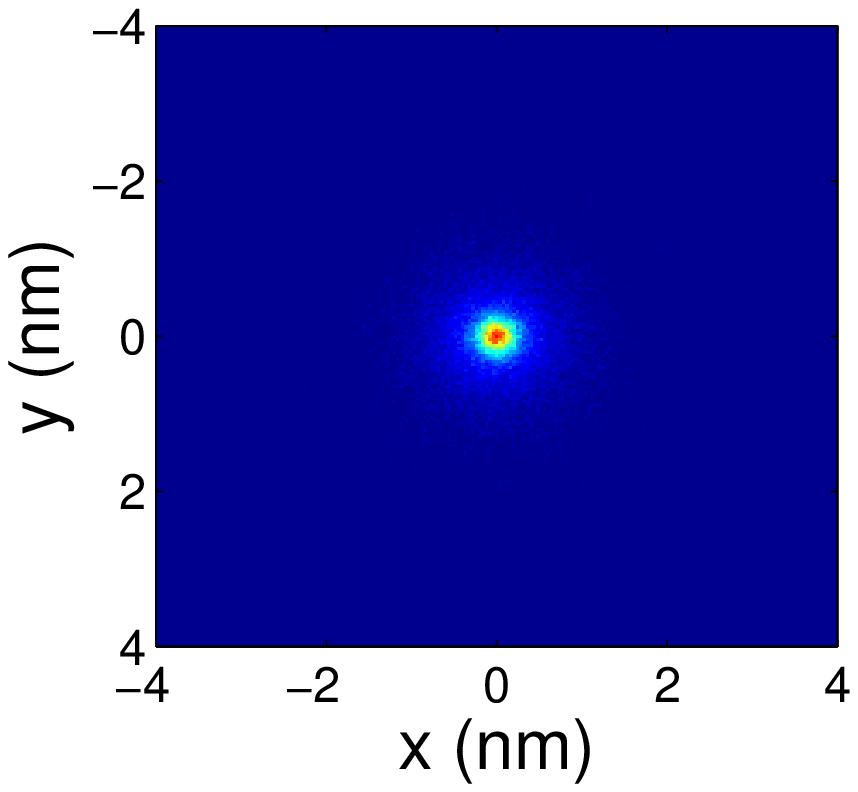}
      \includegraphics[width = 0.22\textwidth]{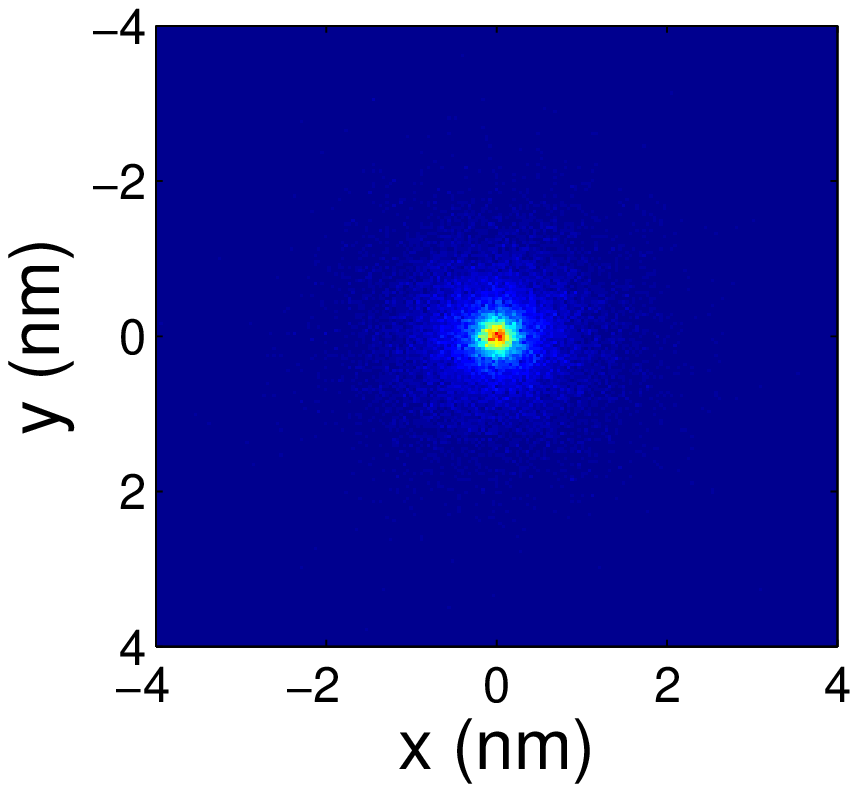}
       \includegraphics[width = 0.22\textwidth]{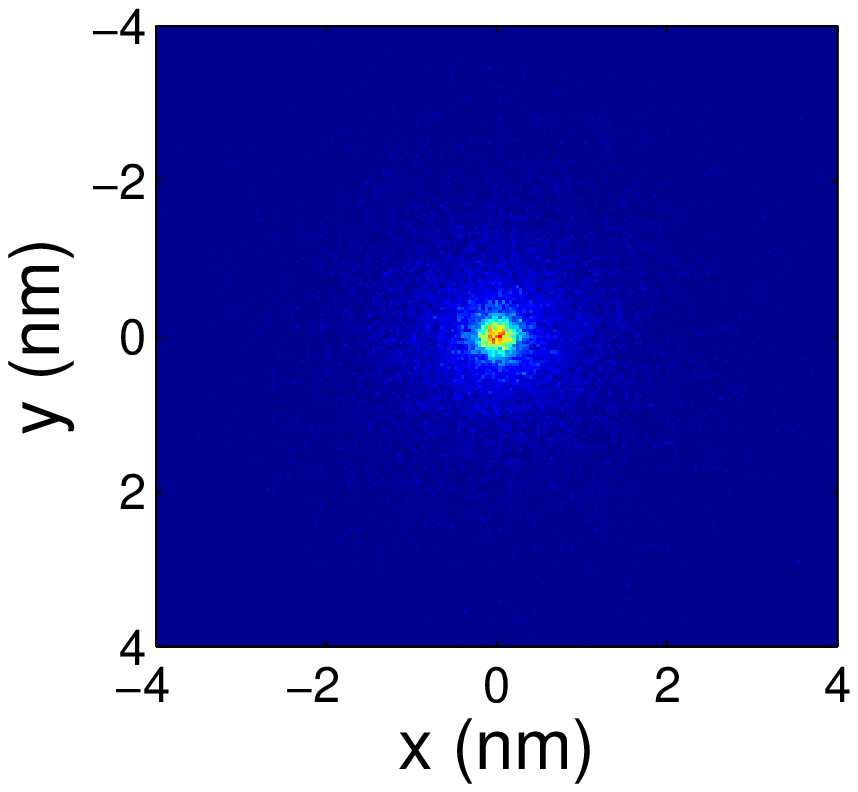}
          \includegraphics[width = 0.22\textwidth]{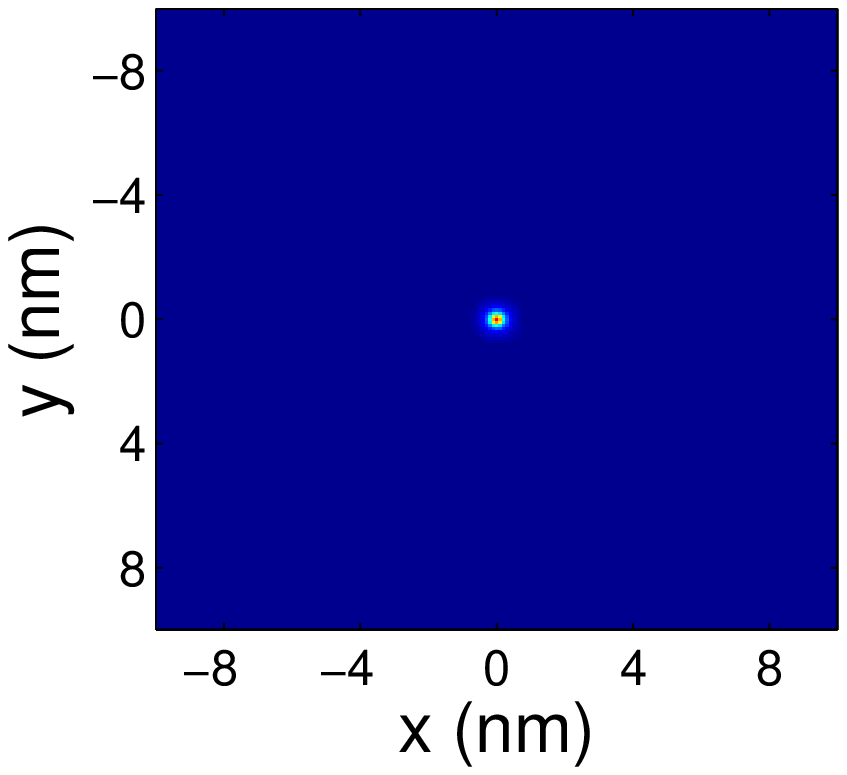}
           \includegraphics[width = 0.22\textwidth]{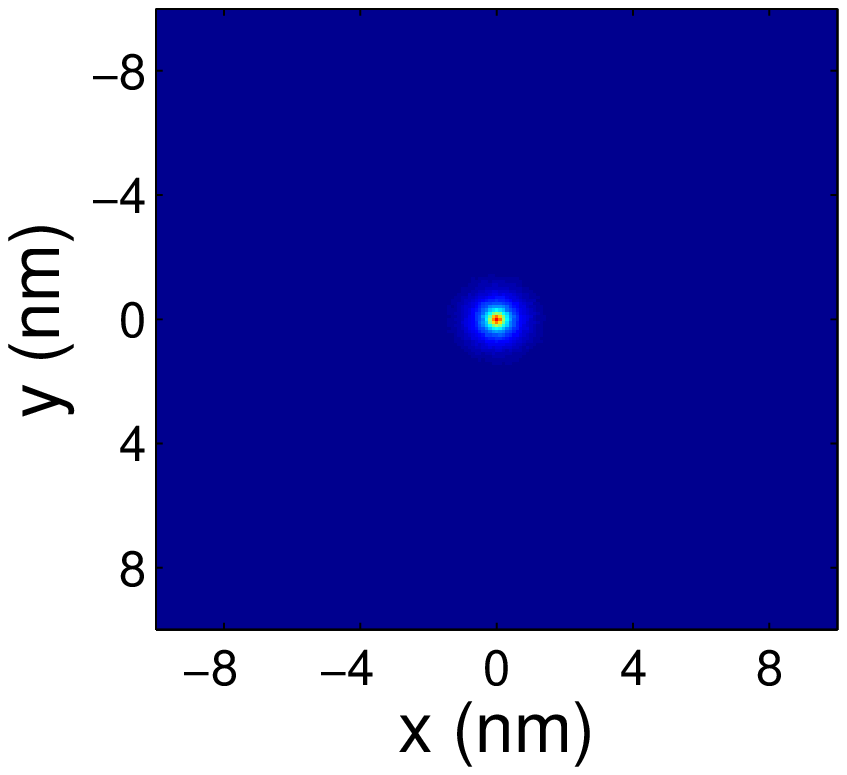}
            \includegraphics[width = 0.22\textwidth]{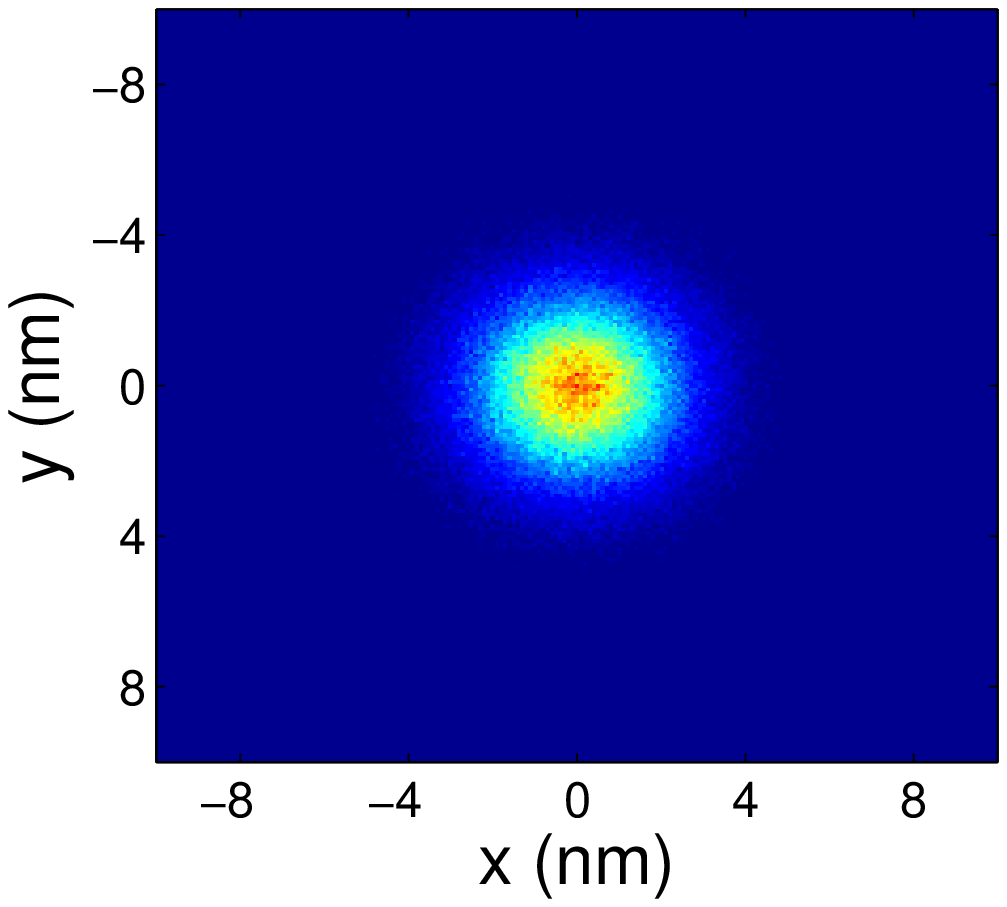}
             \includegraphics[width = 0.22\textwidth]{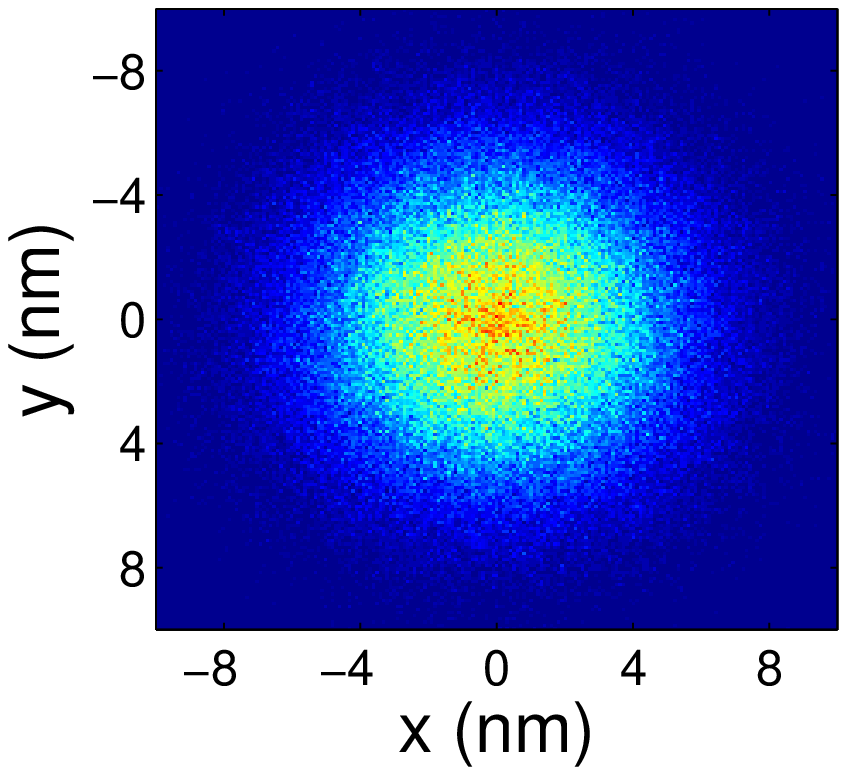}
            \caption{Simulated PSFs for various beam energy spreads (top) and energy offset (bottom). For the top figures, we assume the beam has the optimal energy with energy spread of $10^{-5}$, $5\times10^{-5}$, $10^{-4}$, and $2\times10^{-4}$ from left to right; For the bottom figures we assume the beam energy spread is $10^{-5}$ and the beam energy offset is $10^{-5}$, $2\times10^{-5}$, $5\times10^{-5}$, and $10^{-4}$ from left to right.
    \label{fig1}}
   \end{figure*}

It is worth pointing out that given the same physical length and focal length, the chromatic aberration of a quadrupole magnet is two times smaller than that of a solenoid, considering the fact that the focusing strength of a quadrupole is proportional to $1/B\rho$ and that of a solenoid is proportional to $1/(B\rho)^2$. However, in contrast to a solenoid that focuses beam in both horizontal and vertical directions, a quadrupole focuses beam in one direction while defocuses it in the orthogonal direction. So at least two quadrupoles are needed to form an image in both horizontal and vertical directions and more quadrupoles are needed to make the magnification equal in both directions. Analysis shows that when multiple quadrupoles are used to achieve equal magnification in both planes, the aberrations are, however, different in horizontal and vertical planes. While the aberration in one plane can be made approximately 2 times smaller than that with a solenoid, the aberration in the orthogonal plane will be much larger. Furthermore, imaging with a solenoid is easier from a practical point of view because only the strength of a single solenoid needs to be varied to achieve the imaging condition.

\subsection{Stability requirement}
The stability requirement can be analyzed by taking partial derivative of $R_{12}$ in Eq.~(1), i.e.
\begin{eqnarray}
&&\frac{\partial R_{12}}{\partial L_1}=C-KSL_2=M,
    \nonumber\\
    &&
\frac{\partial R_{12}}{\partial L_2}=C-KSL_1=1/M,
    \nonumber\\
    &&
\frac{\partial R_{12}}{\partial K}=\frac{1}{2K^2(KSL_2-C)}\big[(1-K^2L_2^2)\sin(2KL) \nonumber\\
    &&+2KL_2\cos(2KL) -2K(L+L_2+K^2LL_2^2)\big].
   \end{eqnarray}

The depth of field $D_{ob}$, i.e. the distance along the axis within which the sample can be moved (in practice for example, the position of the sample may have slight variations due to vibration of the ground) without detectable blurring of the image, is correlated with $\partial R_{12}/\partial L_1$. Referred to the object plane, Eq.~(11) indicates that an error of $\Delta L_1$ results in $R_{12}=M\Delta L_1$ that limits the resolution to $R_{12}\sigma'/M=\Delta L_1\sigma'$, where $\sigma'$ is the rms divergence of the beam. Therefore, the depth of field is simply $D_{ob}=\Delta x/\sigma'$, where $\Delta x$ is the resolution of the TEM. For instance, if we aim for 10 nm resolution with a beam with 1 mrad divergence, the depth of field is about 10 $\mu$m. This is also the thickest sample one can image with good focus for both the front and back plane of the sample.

The depth of image $D_{im}$, i.e. the distance along the axis within which the detector can be moved without detectable loss of focus in image, is similarly correlated with $\partial R_{12}/\partial L_2$. Referred to the image plane, Eq.~(11) implies that, compared to $L_1$, the accuracy of $L_2$ is loosened by a factor of $M^2$. So, the depth of image is $D_{ob}=M^2\Delta x/\sigma'$.

Similarly, the sensitivity on variations of solenoid magnetic field and beam energy is correlated with $\partial R_{12}/\partial K$. Assuming the solenoid length is optimized to provide the maximal focusing for a given magnetic field, i.e. ($KL=\pi/2$) and the magnification is much larger than unity, we find approximately $\partial R_{12}/\partial K\approx \pi M/2K^2$. The sensitivity of $R_{12}$ on the relative change of focusing strength (or beam energy) can be characterized by $J=(K/M)\partial R_{12}/\partial K\approx \pi/2K$. Accordingly, the rms size of the image for a point source is
\begin{equation}
\sigma=J\sigma'\Delta B/B,
\end{equation}
where $\Delta B/B$ is the relative offset of the magnetic field with respect to the optimal value (here the beam energy spread is assumed to be negligibly small). This implies that a solenoid with strong strength is required to reduce the sensitivity of an u-TEM on changes of the focusing strength and beam energy. In other words, an u-TEM with stronger solenoid will have a larger momentum acceptance, which is practically useful for good performance in presence of beam energy jitter.

\subsection{Expected performance of a representative u-TEM}
Here we show the performance of a representative u-TEM with a magnification of 10000 using a 3-stage imaging system. The beam energy is assumed to be 4 MeV and the beam rms divergence at the sample is assumed to be 1.5 mrad (comparable to the characteristic scattering angle for Carbon). The focal lengths for the objective lens, intermediate lens and projection lens are 1.25 cm, 2 cm and 2 cm, respectively. The magnetic field for the objective lens is about 2.4 T (corresponding to $K=80$ for $L=2~$cm) and that for both the intermediate lens and projection lens is about 1.5 T. With the magnetic field higher than typical saturation field ($\sim2$~T) of the pole-piece iron of a normal conducting solenoid, a superconducting solenoid (see, for example \cite{SCsolenoid}) should be used for the objective lens. The magnification is 16 for the objective lens and 25 for both the intermediate lens and projection lens. The simulation is performed with ELEGANT code \cite{elegant} that includes the second order effects. The third order effects, e.g. spherical aberration, is not included in the simulation. This is justified because of the small beam divergence and small energy spread.

The simulated point spread functions (PSFs), i.e. the image of a point source for various beam energy spreads and energy offsets (deviation of average beam energy from the optimal value) are shown in Fig.~2. One can see that the effect of beam energy spread on PSF is quite different from that of beam energy offset (see the top-right figure with $10^{-4}$ energy spread and the bottom-right figure with $10^{-4}$ energy offset). For a beam with large energy spread, though a large portion of the particles are out of focus, a small fraction of the particles are still in focus and one main still get a clear image surrounded by a relatively large background. For a beam with small energy spread and large energy offset, however, all the particles are out of focus and the resolution is significantly degraded. Mathematically, this is because deviation of $K$ or $B\rho$ introduces considerable value to $R_{12}$ such that the point-to-point imaging condition no longer holds. In our example with $K=80$ we have $J\approx0.02$, therefore, a relative change of focusing strength (or beam energy) by $10^{-4}$ will lead to a blurring effect of about 3 nm (rms) (calculated by Eq.~(12) with $\sigma'=1.5\times10^{-3}$), in good agreement with the bottom-right figure in Fig.~2.

To show the sensitivity of the image on the electrical stability of the lens power supplies, the focusing strengths of the objective lens and intermediate lens are varied around the optimal values and the FWHM of the PSF is shown in Fig.~3. Recall that the focusing strength is $K=B_0/2B\rho$, changing the strength of the solenoid ($B_0$) is therefore equivalent to changing the average beam energy ($B\rho$). As shown in Fig.~3, to ensure a few nanometer spatial resolution, the variation of the strength for the objective lens should not exceed $10^{-4}$, similar to the requirement of the beam energy offset. For the intermediate lens, as suggested by Eq.~(5), its requirement is significantly loosened (a few nanometer resolution can be achieved with its strength off by 1\%). Similarly, the requirement on stability of the projection lens is even loosened.
   \begin{figure}[h]
       \includegraphics[width = 0.23\textwidth]{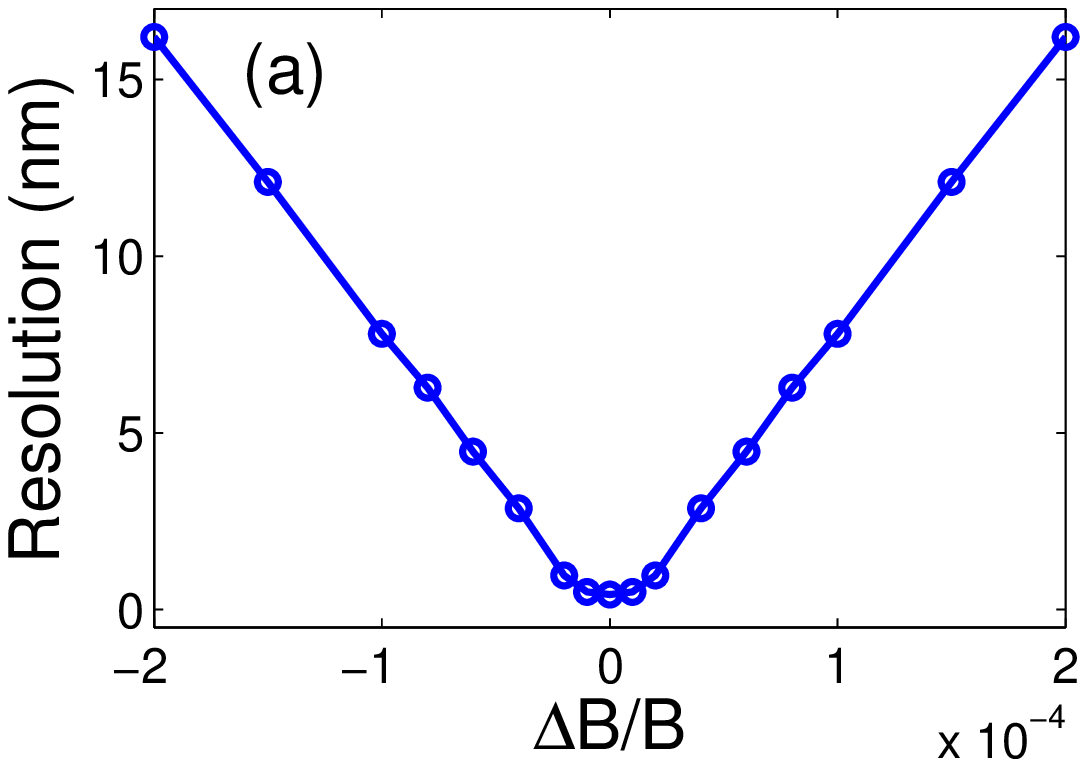}
    \includegraphics[width = 0.23\textwidth]{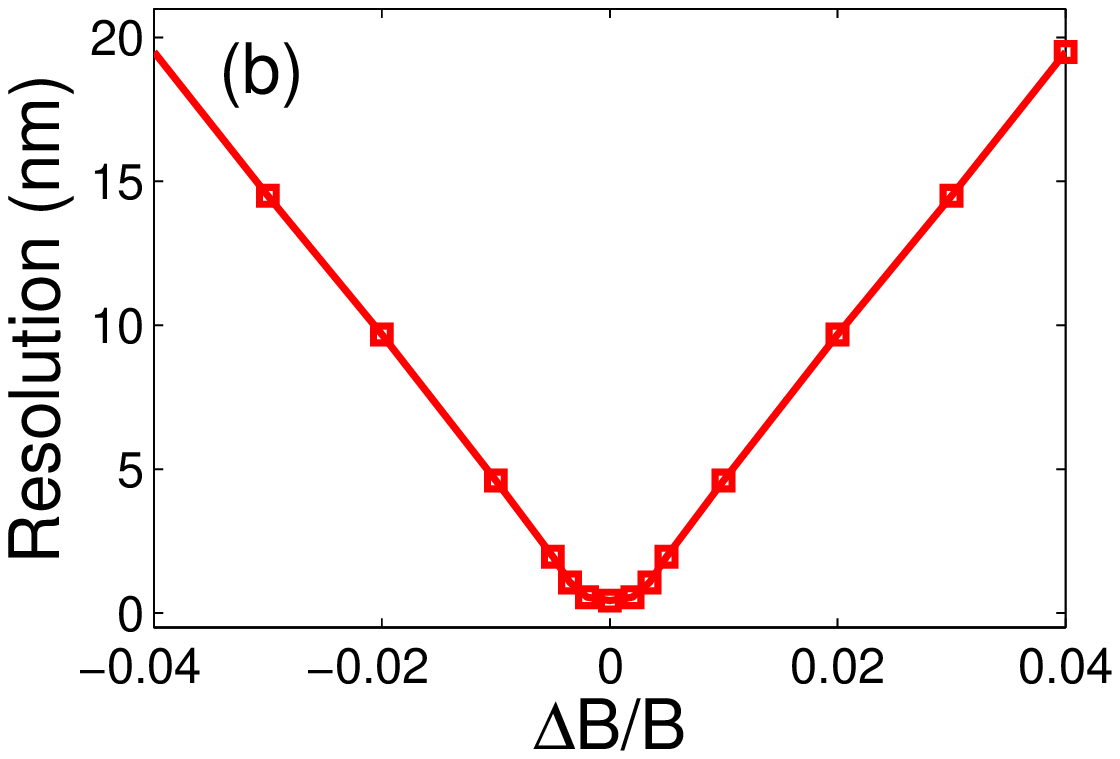}
    \caption{FWHM of the PSF for various deviations of the strengths of the objective lens (a) and the intermediate lens (b).
    \label{beamforuem}}
    \end{figure}

\section{Practical realization of an u-TEM}
Based on the studies in above section, we know the measures that should be taken to ensure good performance of an u-TEM. Specifically, the beam brightness should be made as high as possible by using a high gradient acceleration field. The strength of the objective lens needs to be made as strong as practically possible to reduce the aberrations and to loosen the requirements on beam emittance and energy spread. A good mechanical and electrical stability for the objective lens is also needed. Because the beam divergence is significantly reduced after the objective lens, the stability requirements for subsequent intermediate lens and projection lens are greatly loosened. In this section we show how one can realize an u-TEM with currently available technologies.

\subsection{Electron beam properties from a state-of-the-art photocathode rf gun}
As shown in Fig.~1, for our proposed u-TEM, the electron beam is produced in an S-band photocathode rf gun that is now widely used to provide high brightness electron beam to drive x-ray free-electron lasers (see, for example \cite{LCLS}). Our simulation with the code IMPACT-T \cite{impact} uses the parameters of the LCLS gun with 115 MV/m acceleration gradient \cite{LCLSgun}. In our simulation we assume the drive laser has a flat-top distribution (8 ps full width) in the center and Gaussian ramping (1 ps (rms)) at the head and tail. The laser has a uniform distribution in transverse direction with a diameter of 100 $\mu$m. The initial emittance is assumed to be 10 nm, corresponding to a thermal emittance of 0.4 $\mu$m per mm (rms) laser spot size, similar to the experimental results reported in \cite{PSI}. The simulated beam parameters with 1.6 pC charge are shown in Fig.~4.
  \begin{figure}[h]
       \includegraphics[width = 0.23\textwidth]{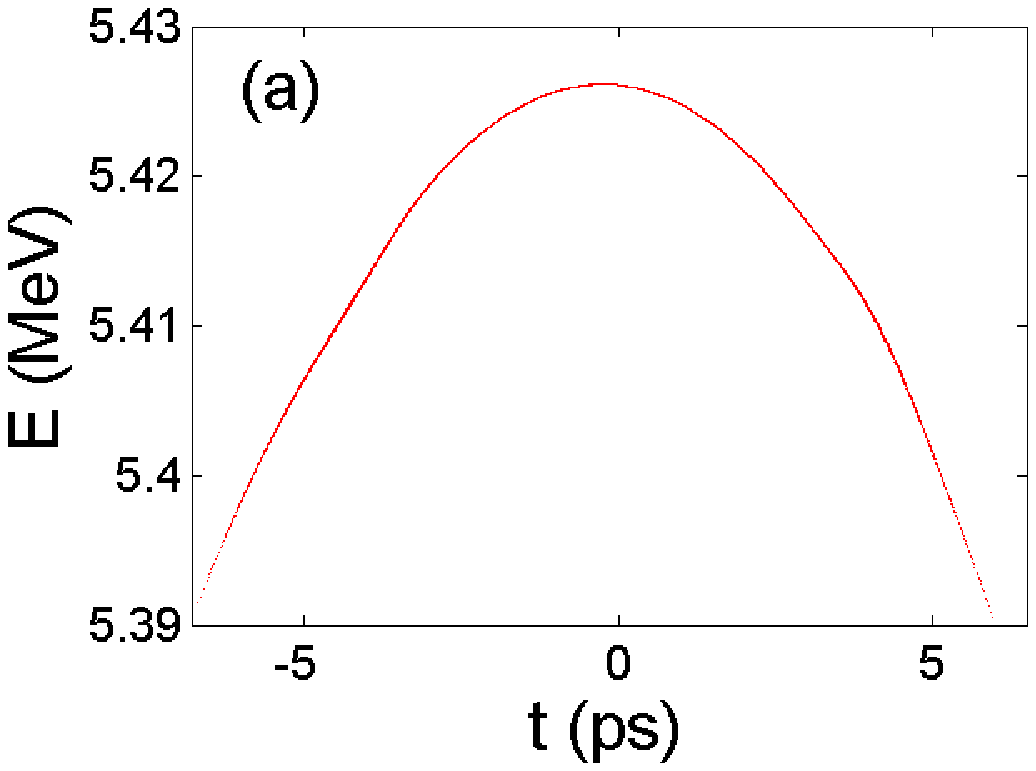}
    \includegraphics[width = 0.23\textwidth]{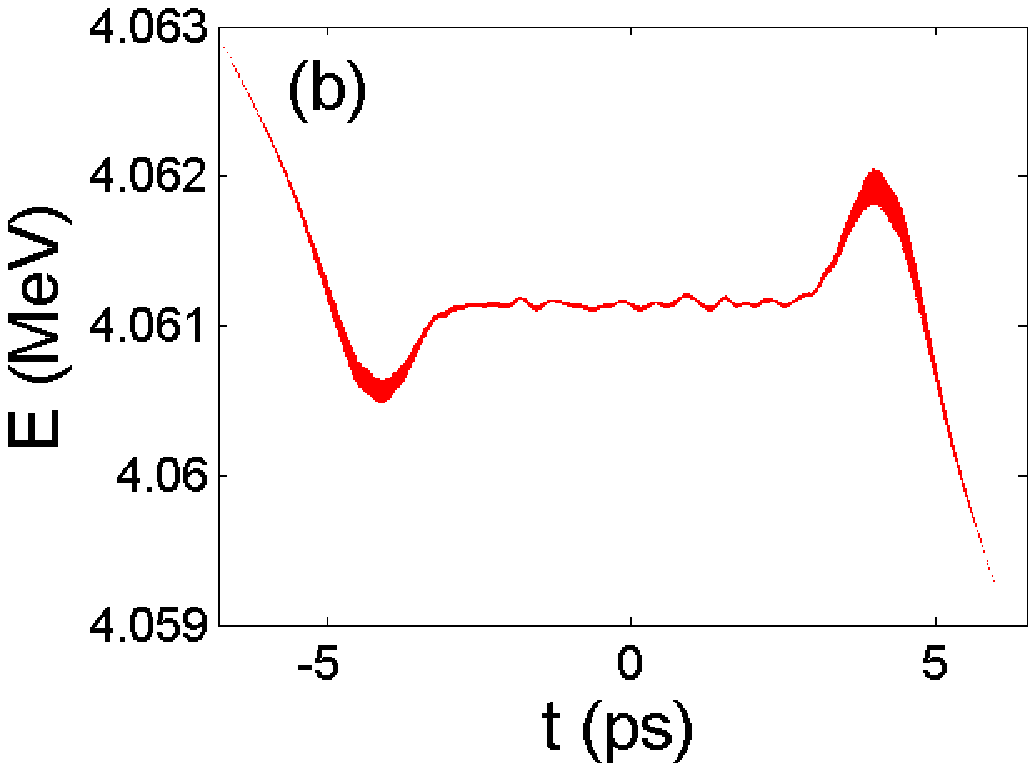}
    \includegraphics[width = 0.23\textwidth]{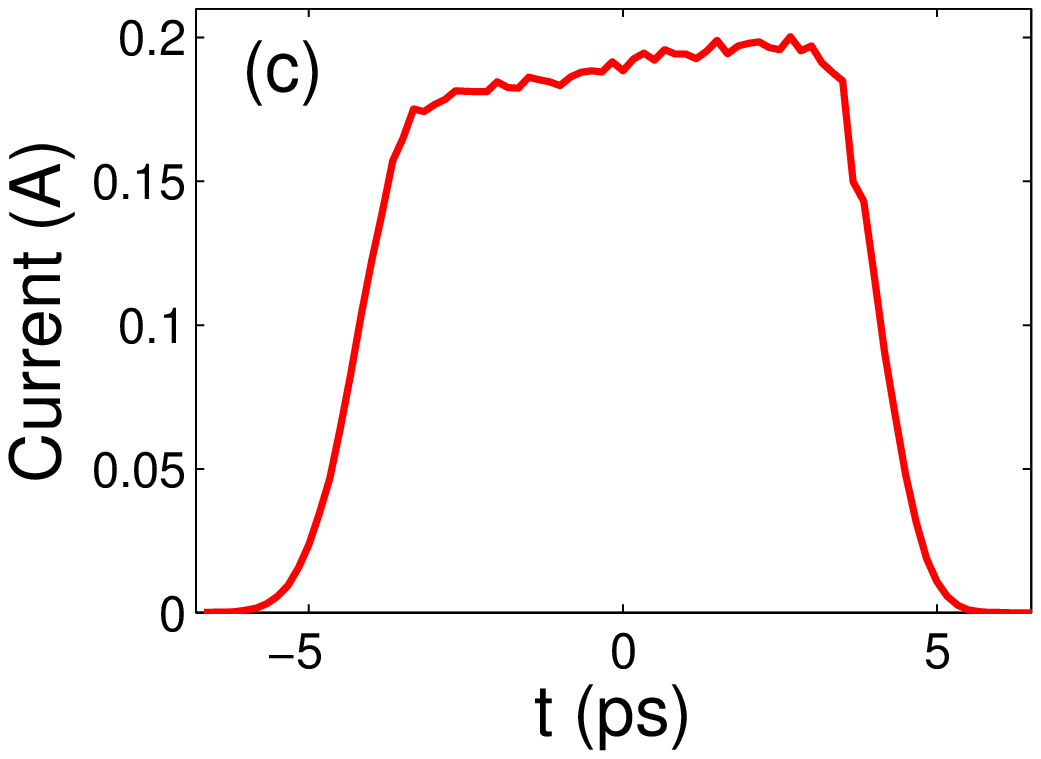}
    \includegraphics[width = 0.23\textwidth]{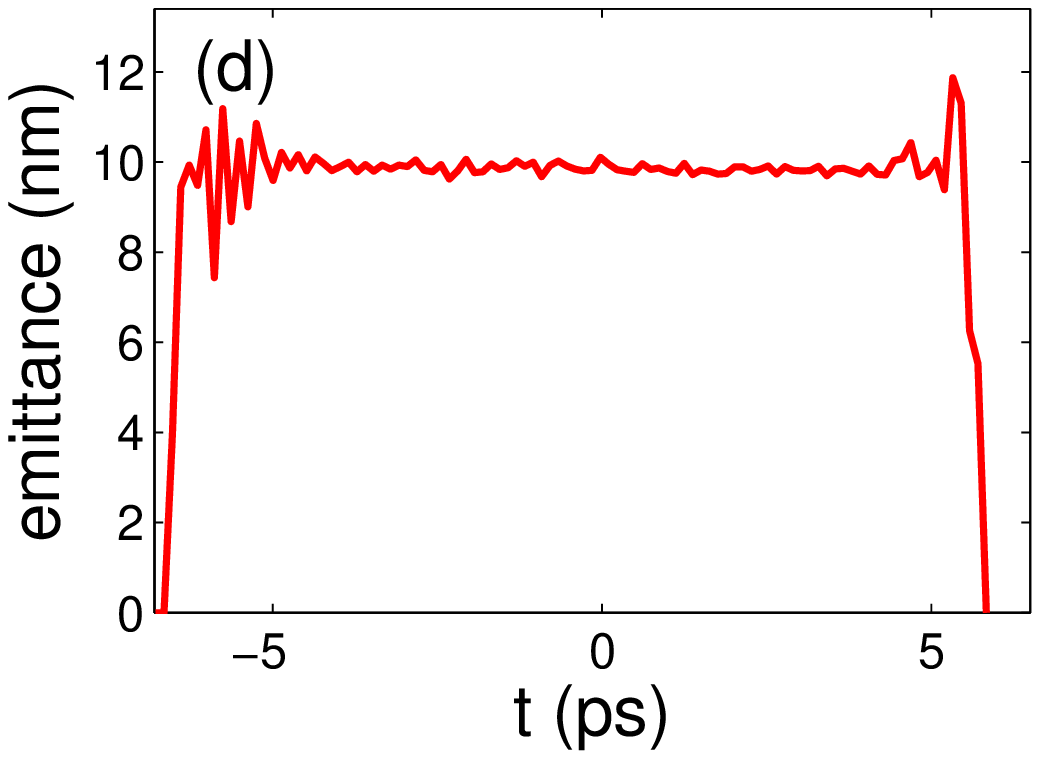}
    \caption{Properties of the electron beam produced in a state-of-the-art photocathode rf gun: (a) beam longitudinal phase space at the exit of the S-band gun; (b) longitudinal phase space after cancelation of the quadratic energy chirp in a C-band cavity; (c) beam current distribution; (d) normalized slice emittance.
    \label{beamforuem}}
    \end{figure}

The beam has a kinetic energy of about 5.4 MeV at the exit of the S-band gun. Because the electron bunch extends over many degrees of the rf phase, the sinusoidal field introduces a correlated spread in beam energy, leading to a global energy spread of about $1\times10^{-3}$ (Fig.~4a). Fortunately, the beam correlated energy spread can be corrected up to second order with an additional rf harmonic cavity. In our simulation, after compensation for the nonlinear rf curvature in a C-band cavity, the global energy spread reduces to $6\times10^{-5}$, and the energy spread for the core beam (within $\pm3~$ps) is only $1\times10^{-5}$ (Fig.~4b). The projected emittance for the whole beam is about 16 nm, with large contributions from bunch head and bunch tail where space charge is strong and nonlinear. The slice emittance as well as the projected emittance for the core beam are both close to the thermal emittance, as shown in Fig.~4d.

To illustrate the physics related to this cancelation, let's assume a beam with energy $E_i$ is accelerated in two rf structures where the main acceleration structure has a wave number at $k_s$ and the harmonic structure used to cancel the quadratic energy chirp has a wave number at $k_c$. The peak energy gain of each structure is assumed to be $E_s$ and $E_c$, and the phase relative to the accelerating peak of the waveform is assumed to be $\phi_s$ and $\phi_c$ ($\phi_s=0$ for obtaining maximal energy gain). The energy of a particle at longitudinal position $z$ (bunch head at $z>0$) with respect to the reference particle can be written as,
\begin{equation}
E(z)=E_i+E_s \cos(\phi_s+k_s z)+E_c \cos(\phi_c+k_c z)
\end{equation}
To obtain maximal energy with effective cancelation of the quadratic chirp, the phase of the main acceleration structure is set to provide maximal energy gain ($\phi_s=0$), and the phase of the harmonic cavity is set at the decelerating phase ($\phi_c=\pi$). Under this condition the relative energy deviation can be written as,
\begin{equation}
\frac{\Delta E(z)}{E_0}=\frac{1}{2}\frac{E_x k_x^2-E_s k_s^2}{E_0}z^2 + ...
\end{equation}
where $E_0=E_i+E_s-E_x$ is the energy of the reference particle. From Eq.~(14) one can see that by properly choosing the voltage of the harmonic cavity, the energy chirp of the beam can be compensated up to the second order. Specifically, the required voltage for the harmonic cavity scales as $1/n^2$ where $n$ is the harmonic number. For instance, when an C-band structure (rf frequency at 5.712 GHz) is used to linearize the curvature developed in an S-band (rf frequency at 2.856 GHz) linac structure, its voltage should be 1/4 of the S-band structure. Figure~4(b) shows effective cancelation of the nonlinear energy chirp for the beam core in a C-band cavity with about 1.36 MV voltage. The head and tail has high order correlations from longitudinal space charge, and can not be compensated for. Note, the beam kinetic energy is reduced to about 4 MeV after deceleration in the C-band cavity.
   \begin{figure}[h]
       \includegraphics[width = 0.23\textwidth]{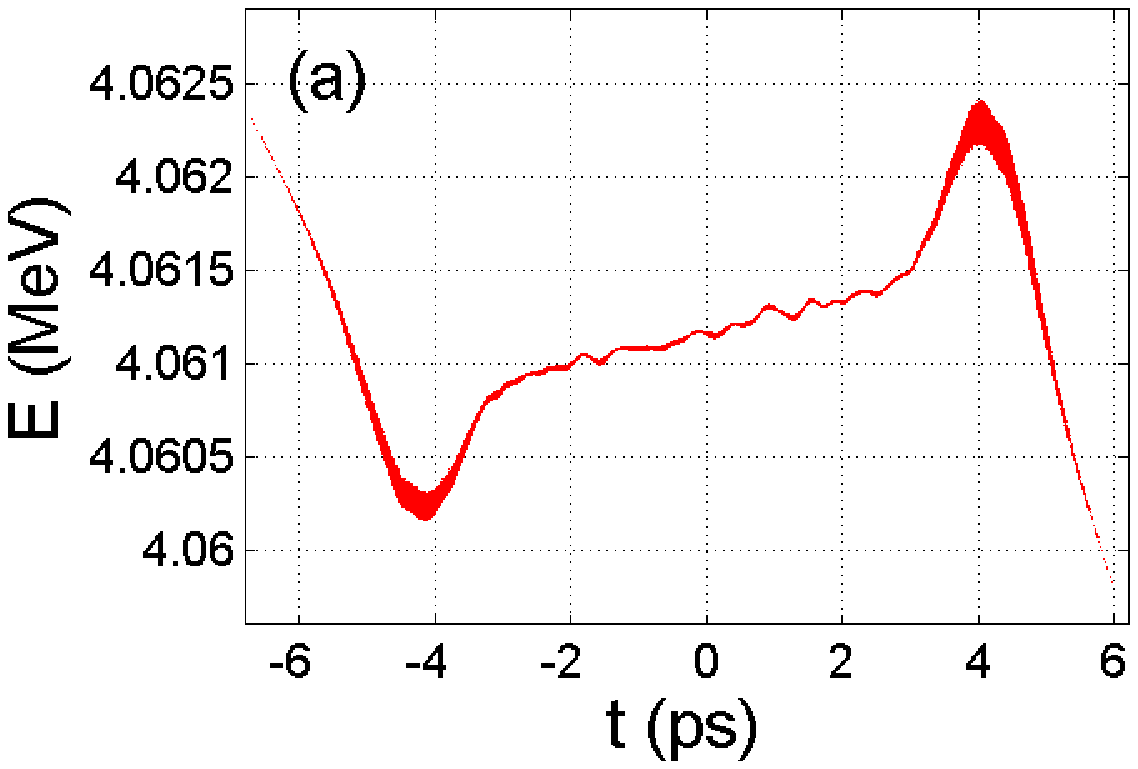}
    \includegraphics[width = 0.23\textwidth]{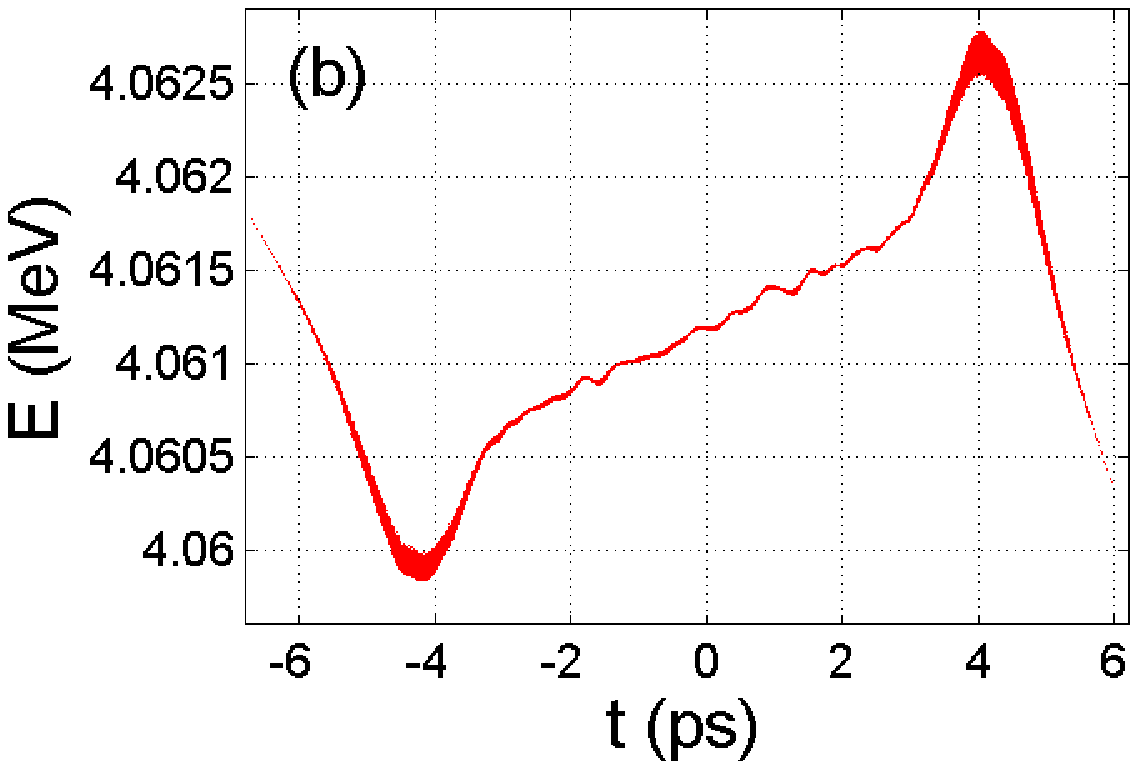}
    \caption{Beam longitudinal phase space at the exit of the C-band cavity with the phase of the cavity off from the optimal value by 0.1 degree (a) and 0.2 degree (b), respectively.
    \label{beamforuem}}
    \end{figure}

It should be pointed out that for cancelation of second order chirp, $\phi_s$ and $\phi_c$ don't necessarily need to be 0 and $\pi$, respectively. A small deviation from these values still provides effective cancelation of the second order chirp (because $\cos\phi\approx1$ when $\phi\ll1$), but it will lead to a first order energy chirp as $\Delta E(z)=-E_ck_c(\sin\phi_s+\sin\phi_c)z$. While in theory for a given $\phi_s$, one can always find the right $\phi_c$ to zero the first order chirp, in practice the rf phase jitter makes this cancelation very difficult. For instance, with the phase of the C-band cavity off from the optimal value by 0.1 degree, a linear chirp is seen in the beam longitudinal phase space that leads to an energy variation of about $1.4\times10^{-4}$ for the core of the electron beam (Fig.~5a). A 0.2 degrees phase difference similarly leads to an energy variation of about $2.8\times10^{-4}$ in the beam core (Fig.~5b). So the phase jitter of the rf systems needs to be controlled to the level of 0.1 degree in order to obtain a beam with energy spread on the order of $10^{-4}$. Note, the variations of $E_s$ and $E_c$ lead to a change of the average beam energy and therefore the amplitude jitter of the rf system needs to be controlled to better than $10^{-4}$ to realize a few nanometer spatial resolution, as illustrated in Fig.~2 and Fig.~3.
    \begin{figure*}[t]
    \includegraphics[width = 0.325\textwidth]{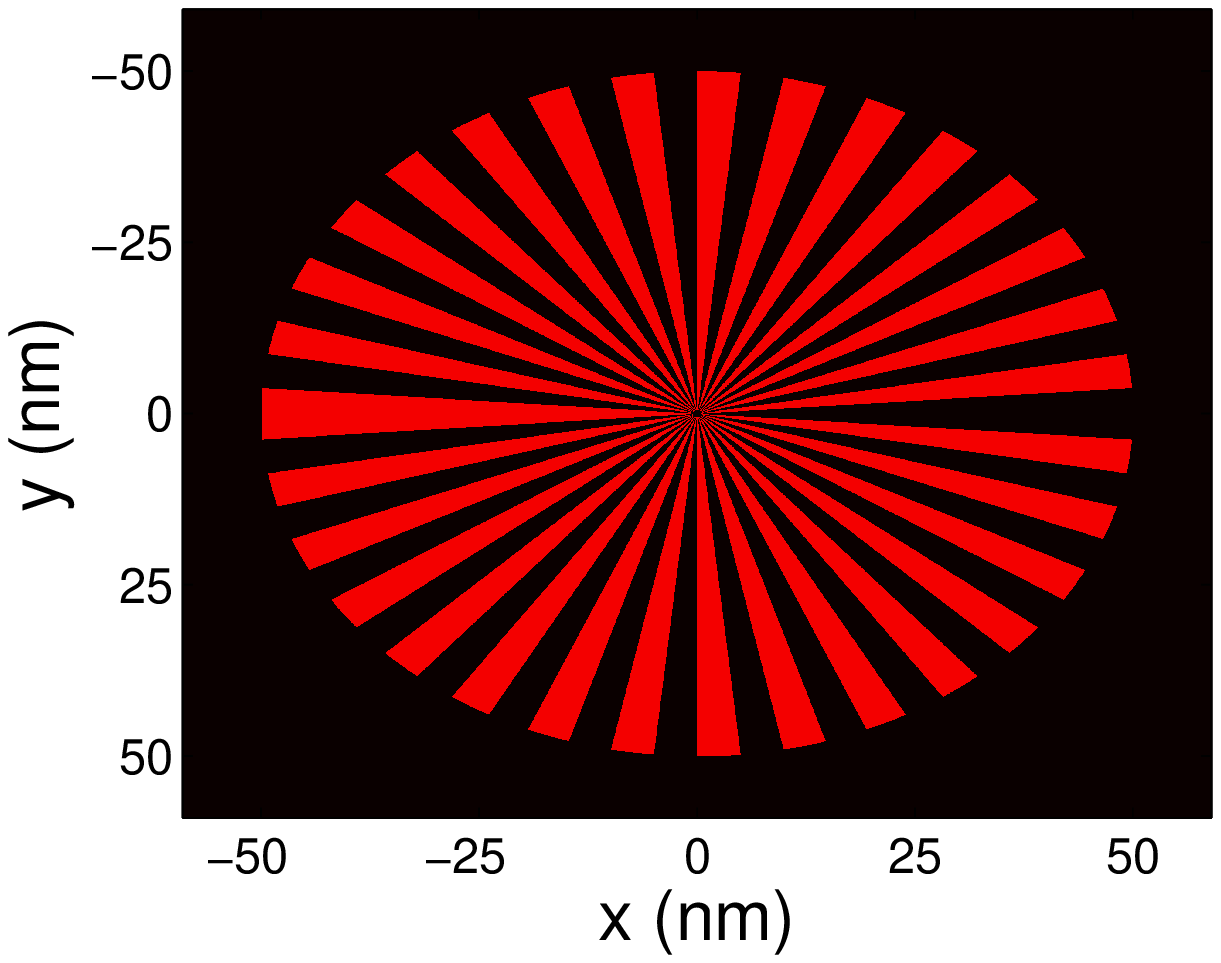}
   \includegraphics[width = 0.325\textwidth]{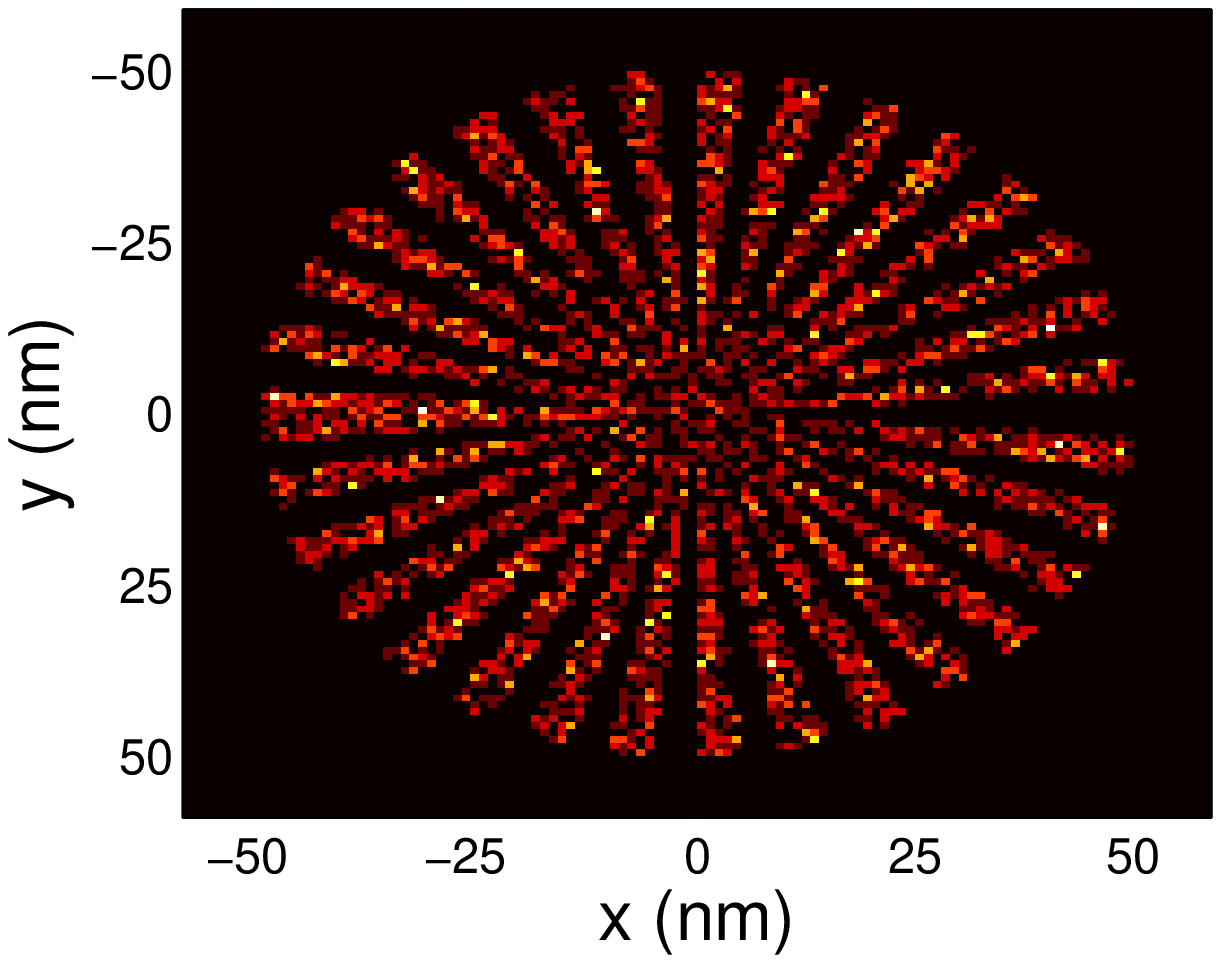}
   \includegraphics[width = 0.325\textwidth]{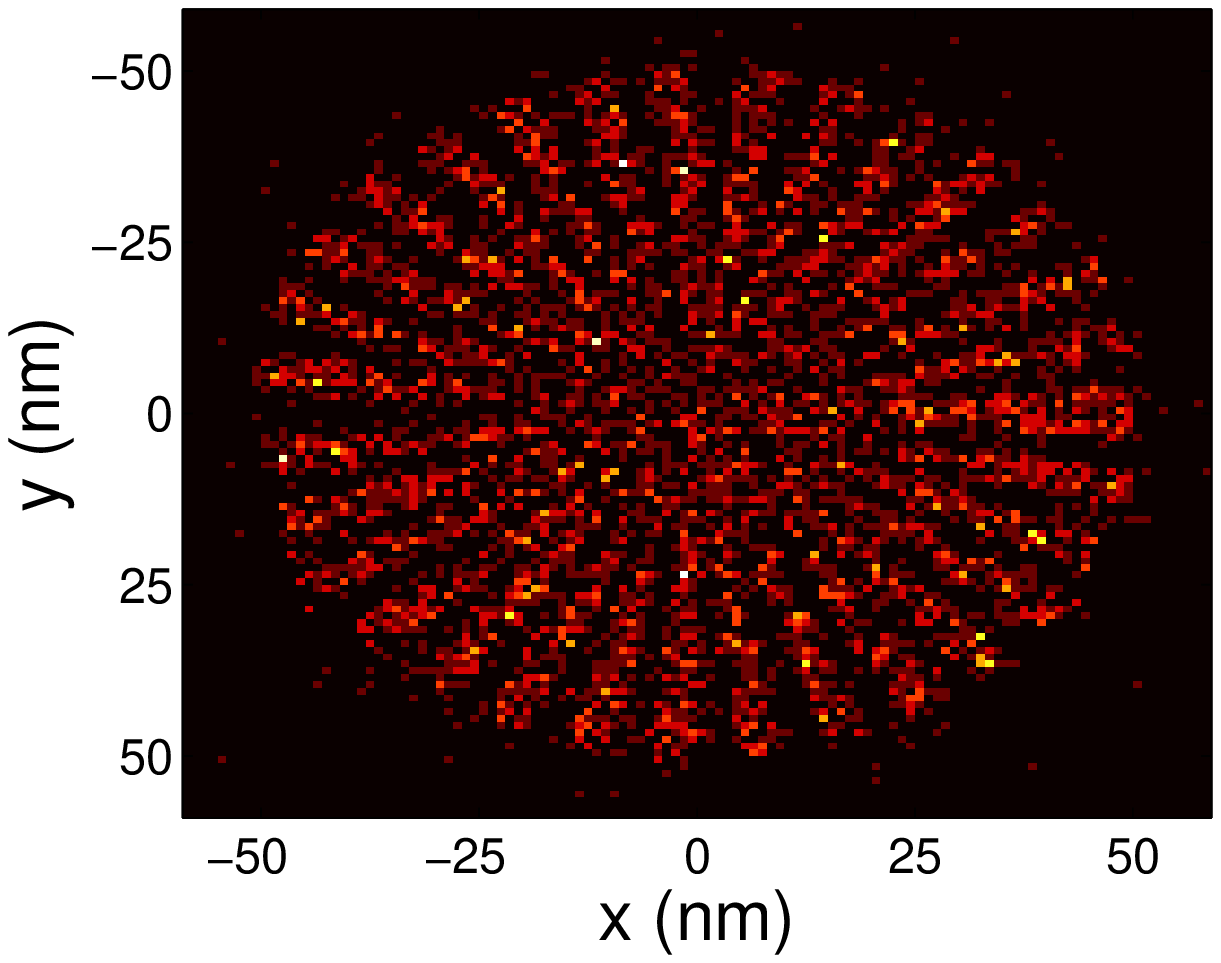}
            \caption{Distribution of the virtual sample (left), beam distribution at the virtual sample (middle) and the image (right) with 1 nm binning. Note, the image is inverted.
    \label{fig1}}
   \end{figure*}

\subsection{Expected performance with a practical beam}
In this section, we present the performance of a representative u-TEM using the realistic beam distributions (as shown in Fig.~4) with space charge effect taken into account using the code PARMELA \cite{PARMELA}. At least two effects from space charge force are worthy of comment. First, the longitudinal space charge introduces an almost linear chirp in the beam center (beam energy varies by about $10^{-4}$ for the beam core) during the transportation to the sample where the beam is focused down to 1 $\mu$m with the condenser lens. While this may be pre-compensated by running the C-band cavity at a phase slightly different from the optimal value to imprint a small chirp that together with the space charge induced chirp leads to minimized energy spread at the sample, in our simulation the space charge induced energy chirp was not corrected. Considering the fact that the relative phase between the S-band gun and the C-band cavity may have fluctuations on the order of a fraction of a degree, our simulation with the space charge induced energy chirp uncompensated is a more accurate representation of the practical case. Second, the transverse space charge introduce a considerable defocusing to the beam, and in the simulation the strength of the objective lens is increased by about $2.5\times10^{-4}$ (compared to the optimal value without considering space charge effect) to compensate for the defocusing effect. The strengths of the intermediate lens and projection lens are kept the same as those used in section above, because they have quite loose tolerance from the reduced beam divergence.

As a representative example, in the simulation we assumed an ideal 100\%-contrast sample as shown in the left plot of Fig.~6 (electrons striking on black regions are all ``absorbed'' by the sample). The periodicity of the tapered structure is 10 nm at the outer edge of the sample (50 nm away from the center) and linearly decreases as it approaches the center. The beam distribution immediately after the sample and the image formed at the detector referred back to the sample plane (i.e. with the image demagnified by 10000) are shown in the middle and right plot of Fig.~6, respectively. This image is formed with about 5200 detected electrons ($10^7$ electrons corresponding to 1.6 pC charge are used in the simulation) with 1 nm binning. This corresponds to a measurement with a screen with 10 microns resolution. The highest spatial frequency (above noise level) of the image is quantified from a 2-D Fourier transformation, and the resolution is found to be approximately 4 nm.

It should be pointed out that on average there are only about 3 electrons per pixel in Fig.~6. So in practice the image shown in Fig.~6 may require integration over $\sim$10 electron pulses. Nevertheless, with a pixel size of about 50 microns (corresponding to 5 nm resolution), there may be sufficient signal for obtaining a useful image in a single-shot.

In Fig.~7 we plot the beam distribution at the image plane for the core part of the beam and the head/tail part of the beam. The beam head and tail that contain about 30\% of the particles with a relatively large energy spreads and energy offsets only produce a background that reduces the image contrast (right plot in Fig.~7). Image formed with the core beam (left plot in Fig.~7) shows improved resolution and contrast compared to that formed with the whole beam (right plot in Fig.~6). This suggests that the beam may be energy-filtered to enhance the performance of an u-TEM.
   \begin{figure}[h]
       \includegraphics[width = 0.23\textwidth]{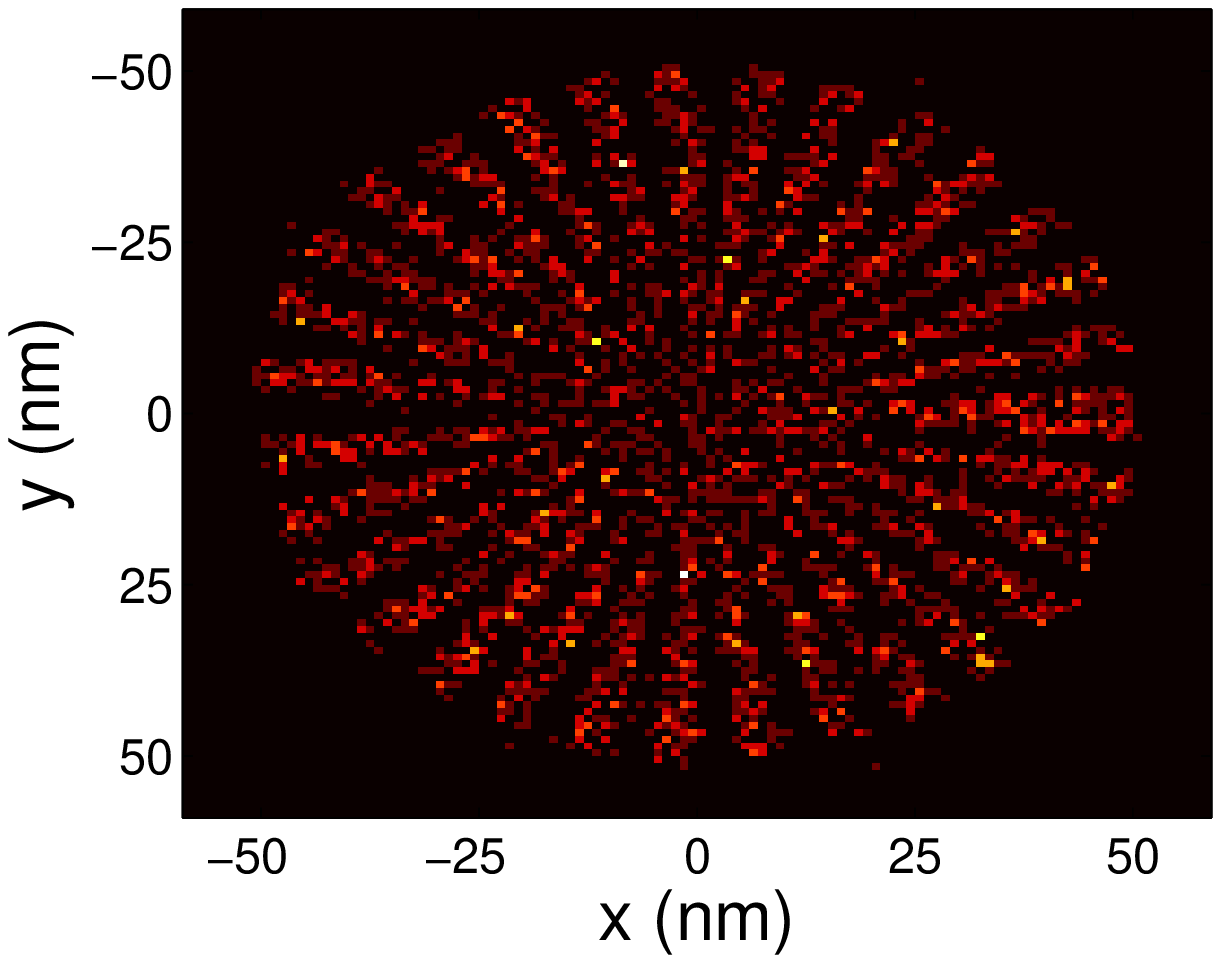}
    \includegraphics[width = 0.23\textwidth]{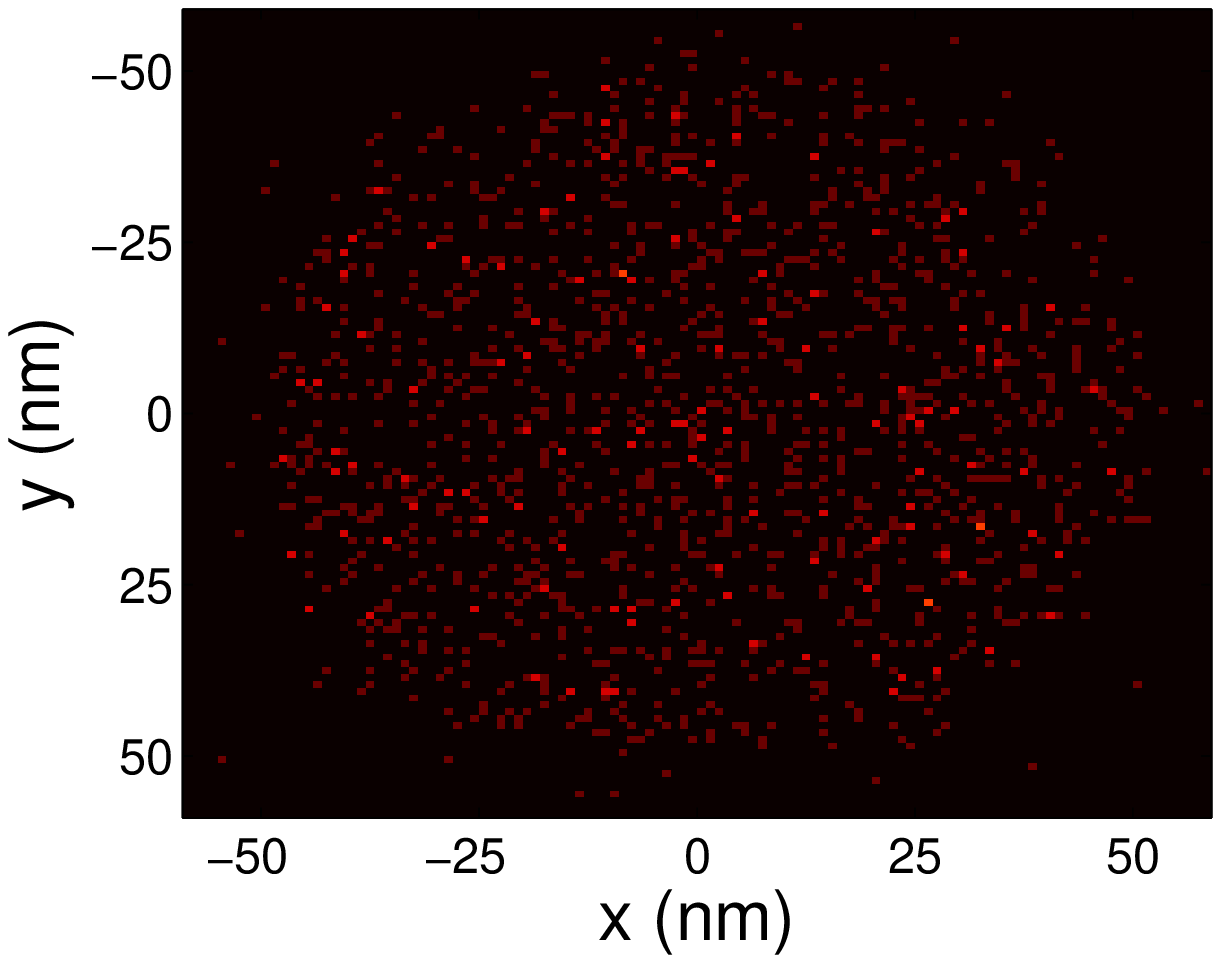}
    \caption{Image distribution formed with the beam core (left) and with the head and tail of the beam (right).
    \label{beamforuem}}
    \end{figure}

\section{Summary and discussions}
Using realistic beam parameters that can be obtained with a state-of-the-art photocathode rf gun, we have shown that an u-TEM with a few picosecond time resolution and a few nanometer spatial resolution is feasible. The most challenging requirement seems to be the $10^{-4}$ (peak-to-peak) energy stability (for objective lens with $2\sim3$~T magnetic field), which has not yet been demonstrated (e.g. the rms energy jitter of LCLS gun is about $2\times10^{-4}$ \cite{LCLSgun}). Possible solutions include development of superconducting rf gun that may provide a beam with $10^{-5}$ energy stability, using a higher field superconducting solenoid or correcting the aberrations (see, e.g. Ref.~\cite{firstcorrection,50pm}) to loosen the requirement on beam energy stability, etc.

It is worth pointing out that the parameters used in our simulations and calculations are representative rather than fully optimized design sets. A more careful optimization should lead to further improvements of u-TEM. It should be pointed out that in order to obtain analytical formula and simplify the analysis, we used a hard-edge model for the solenoid. Furthermore, only the global space charge force is included in our analysis while the more complicated Coulomb collisions from the nearest neighboring particles are neglected. A more complete study with realistic bell-shaped field for solenoids and with Coulomb collision effects included will be performed in the future.

\section{Acknowledgements}
We thank Alex Chao and Renkai Li for useful discussions. This work was supported by the National Natural Science Foundation of China under contract No. 11327902, and by the U.S. DOE under Contract Nos. DE-AC02-76SF00515 and DEAC02-05CH11231. One of the authors (DX) wants to thank the support from DOE Early Career Award from 2012 to 2014.

\section{Appendix}
Here we briefly derive the second order transfer matrix for a hard-edge solenoid with Lie algebra. In the usual canonical coordinates ($x$,$p_x$,$y$,$p_y$,$z$,$\delta$), to third order the Hamiltonian that describes the beam motion in a solenoid field is $H = H_2 + H_3$, with
\begin{eqnarray}\label{eqH23}
H_2 &= \frac12 ((p_x+ky)^2+(p_y-kx)^2+\delta^2/\gamma_s^2), \nonumber \\
H_3 &= -\frac12 \delta ((p_x+ky)^2+(p_y-kx)^2+\delta^2/\gamma_s^2),
\end{eqnarray}
where $\gamma_s$ is the Lorentz factor of the reference particle (see, e.g.~\cite{DragtSolenoid,IselinMADPhy}). In Eq. (\ref{eqH23}) we have assumed the reference particle is relativistic with $\beta_s\approx1$.

For an accelerator element whose Hamiltonian is not $s$-dependent ($s=\beta_s ct$ is the free variable), the Lie map is
\begin{eqnarray}
X_f &=& e^{:f:}X_i, \quad f=-H L,
\end{eqnarray}
where $L$ is the length of the element, $f$ is the generating function of the Lie map, $X_i$ and $X_f$ are 6D coordinate vectors at the entrance and exit planes, respectively, and the operator $:f:$ signifies the Poisson bracket,
i.e.,
\begin{equation}
:f:g = [f,g]=\sum_{i=1}^3 \frac{\partial f}{\partial q_i}\frac{\partial
    g}{ \partial p_i} -\frac{\partial f}{\partial p_i}\frac{\partial
    g}{\partial q_i}
\end{equation}
for a function $g$, with $p_i$, $q_i$, $i$=1,2,3 being the canonical coordinates.
The Lie map can be decomposed into a series of linear and nonlinear maps
\begin{eqnarray}
e^{:f:}&=& e^{:f_2:}e^{:f_3:} \cdots,
\end{eqnarray}
where $f_2$ and $f_3$ are second and third order homogeneous polynomials, respectively.

\subsection{First order map for a solenoid}
The first order transfer matrix of an accelerator element can be derived from the second order terms of the Hamiltonian of motion, $H_2$. The second order generating function of the Lie map is
\begin{eqnarray}
f_2=-H_2 L=-\frac 12 \tilde{X} FX,
\end{eqnarray}
where $F$ is a $6\times 6$ symmetric matrix and $ \tilde{X}$ is the transpose of the 6D coordinate vector $X$. Defining the asymmetric matrix
\begin{eqnarray}
S=\left( \begin{array}{ccc} S_2 & 0 & 0 \\ 0 & S_2 & 0 \\ 0 & 0 & S_2 \end{array} \right),
\qquad {\rm with} \quad
S_2=\left( \begin{array}{cc} 0 & 1 \\ -1 & 0 \end{array} \right),
\end{eqnarray}
the first order transfer matrix $R$ is then given by~\cite{ChaoNotesLie}
\begin{eqnarray}\label{eqRtransMat}
R &=& e^{SF}.
\end{eqnarray}
Using Eqs.~(\ref{eqH23}) and (\ref{eqRtransMat}) we can find the first order transfer matrix of a solenoid as given in Ref. ~\cite{handbook, IselinMADPhy}.

\subsection{ Second order map for a solenoid}
The second order map can be derived from the first order transfer matrix $R$ and the third order terms of the Hamiltonian of motion $H_3$. The third order generating function of the Lie map is~\cite{ChaoNotesLie}
\begin{eqnarray}
f_3=-L\int_0^1  e^{:uLH_2:} H_3 du,
\end{eqnarray}
which is a third order homogenous polynomial
\begin{eqnarray}
f_3=C_{lmn}X_l X_m X_n,
\end{eqnarray}
where $C_{lmn}$, $l,m,n=1,2,\cdots,6$ are the coefficients, $X_{l,m,n}$ are components of the canonical coordinate vector $X$
and summation is assumed for $l$,$m$ and $n$ whenever a pair of identical indices appear in the same term. The elements of the second order map is then given by~\cite{ChaoNotesLie}
\begin{eqnarray}\label{eqT2map}
T_{ijk} &=& -3 S_{i l} C_{l mn} R_{mj}R_{nk}.
\end{eqnarray}
The same summation convention applies in Eq. (\ref{eqT2map}).

Knowing the transfer matrix and the third order generating function, all second order transport map elements can be readily calculated. In particular, the $T_{126}$ element for a hard-edge solenoid is given by
\begin{eqnarray}\label{eqT126}
T_{126} &=& -\frac{L}2 \cos 2kL.
\end{eqnarray}

\subsection{Decoupled transfer map in ($x$,$x'$,$y$,$y'$,$z$,$\delta$) coordinates}
In previous sections the results are given for ($x$,$p_x$,$y$,$p_y$,$z$,$\delta$) coordinates.
Some tracking codes such as ELEGANT~\cite{elegant} as well as the analysis in this paper use ($x$,$x'$,$y$,$y'$,$z$,$\delta$) coordinates, where
the coordinates $x'$, $y'$ are related to $p_x$, $p_y$ through (to first order)
\begin{eqnarray}\label{eqPxPyxpyp}
p_x &=& x' (1+\delta), \quad p_y=y' (1+\delta).
\end{eqnarray}
The map elements given in ($x$,$p_x$,$y$,$p_y$,$z$,$\delta$) coordinates can be readily converted to
elements in  ($x$,$x'$,$y$,$y'$,$z$,$\delta$) coordinates with Eq. (\ref{eqPxPyxpyp}). For example,
\begin{eqnarray}\label{eqT126bar}
\bar{T}_{126} &\equiv& \frac12 \frac{\partial^2 x_f}{\partial x' \partial \delta} =-\frac{L}2 \cos(2kL)+ \frac{\sin 2kL}{4k}.
\end{eqnarray}
Similar results can be found for other elements that involve $x'$ and $y'$.

Finally, we rotate the coordinate by the angle $\phi = -k L$ to decouple the transfer matrix of a solenoid between the two transverse directions. The coordinate transformation can be seen as the application of a transfer matrix
\begin{eqnarray}\label{eqRot}
X^{(2)}_i &= R^{\rm rot}_{ij} X^{(1)}_j,
\end{eqnarray}
where $X^{(1)}$ and $X^{(2)}$ represent coordinates before and after rotation, respectively. Inserting
the Taylor map
\begin{eqnarray}\label{eqTaylor}
X^{(1)}_j &= R_{jk} X^{(0)}_k+T_{jkl}X^{(0)}_kX^{(0)}_l+O(3),
\end{eqnarray}
to Eq. (\ref{eqRot}), we obtain
\begin{eqnarray}\label{eqTaylor2}
X^{(2)}_i &=& R^{\rm rot}_{ij} R_{jk} X^{(0)}_k+R^{\rm rot}_{ij} T_{jkl}X^{(0)}_kX^{(0)}_l+O(3), \nonumber \\
  &=& R^{\rm new}_{ik} X^{(0)}_k+T^{\rm new}_{ikl}X^{(0)}_kX^{(0)}_l+O(3),
\end{eqnarray}
from which we identify the new transfer matrix $R^{\rm new}$ and second order map $T^{\rm new}$,
\begin{eqnarray}\label{eqMapNew}
R^{\rm new} &=& R^{\rm rot} R, \\
T^{\rm new} &=& R^{\rm rot} T.
\end{eqnarray}
With Eq.~(\ref{eqMapNew}), the decoupled transfer matrix in ($x$,$p_x$,$y$,$p_y$,$z$,$\delta$) coordinates can be found, and then using Eq. (\ref{eqT126bar}) one obtains the second order decoupled transfer matrix in ($x$,$x'$,$y$,$y'$,$z$,$\delta$) coordinates. For instance, the $T_{126}$ element of a solenoid as given in Eq.~(9) with the rotated coordinates is found to be,
\begin{eqnarray}
{T}^{(s)}_{126} &=& -\frac{L}2 \cos kL +  \frac{\sin kL}{2k}.
\end{eqnarray}
Other second order transfer matrix elements can be found in a similar way.

\end{document}